\documentstyle[11pt]{article}
\newwrite\ffile\global\newcount\figno \global\figno=1

\def\writedef#1{}

\title{Black hole hair: twenty--five years after}

\author{Jacob D. Bekenstein\\
Racah Institute of Physics, The Hebrew University\\
Givat Ram, Jerusalem 91904 ISRAEL\\
bekenste@vms.huji.ac.il}
\begin{document}
\setlength{\baselineskip 12pt}
\maketitle

\begin{abstract}
Originally regarded as forbidden, black hole ``hair'' are fields associated
with a stationary black hole apart from the gravitational and electromagnetic
ones. Several stable stationary black hole solutions with gauge or Skyrme
field hair are known today within general relativity.  We formulate here a
``no scalar--hair'' conjecture, and adduce some new theorems that almost
establish it by ruling out - for all but a small parameter range - scalar
field hair of spherical black holes in general relativity, whether the field be
self--interacting, coupled to an Abelian gauge field, or nonminimally coupled
to gravity.
\end{abstract}

Twenty--five years ago Wheeler enunciated the Israel--Carter conjecture,
today colloquially known as ``black holes have no hair'' \cite{Wheeler}.
Inmensely influential in black hole physics, this conjecture has long been
regarded as a theorem by large sectors of the gravity--particle physics
community.   On the other hand, the proliferation in the 1990's of solutions
for stationary black holes equipped with various exterior fields
\cite{gauge} has led many modern workers to regard the conjecture as
having fallen by the wayside.   As I hope to make clear here, both these
extreme views are off the mark.

How did this complicated situation emerge ?  By 1965 the charged
Kerr--Newman black hole metric was known.  Inspired by Israel's
uniqueness theorems for the Schwarzschild and Reissner--Nordstr\"om (RN) black
holes \cite{Israel} and Carter and Wald's uniqueness theorems for the
Kerr black hole \cite{Carter},  Wheeler anticipated that ``collapse leads to
a black hole endowed with mass and charge and angular momentum, but, so far as
we can now judge, no other free parameters".  Wheeler stressed that quantum
numbers such as baryon number or strangeness can have no place in the
external description of a black hole.

Wheeler focused on mass, electric charge and angular momentum because they
are all conserved quantities subject to a Gauss law. One can
thus determine these properties of a black hole by measurements from afar.
Obviously this reasoning has to be completed by including magnetic
charge as a fourth parameter; magnetic charge is conserved in
Einstein--Maxwell theory, posesses a Gauss law, and duality of the theory
permits Kerr-Newman like solutions with magnetic charge alongside electric
charge.  We could define ``hair'' as any field not of gravitational or
electromagnetic nature associated with a black hole.  Alternatively, since
there are now many black hole solutions with gauge fields (which are not that
different from Maxwell's field), one may define hair as those free {\it
parameters\/} of the black hole which are not subject to a Gauss law.

Why is ``hair'' so interesting ?   Black holes in their role as gravitational
solitons are to nonperturbative gravity theory as atoms are to quantum theory
of matter and chemistry. Black hole mass and charge are analogous to atomic
mass and atomic number.  Thus if black holes can have other parameters, such
hairy black holes  would be analogous to excited atoms and radicals, the stuff
of exotic chemistry.   Additionally, the absence of a large number of hair
parameters, which is still a viable option today,  is support for the
conception of black hole entropy as the measure of the vast number of internal
(microscopic) degrees of freedom.  Indeed, the ``no--hair'' conjecture
inspired the formulation of black hole thermodynamics
\cite{BekPT}.

Originally ``no--hair theorems'' meant theorems of uniqueness of the
Kerr--Newman family within the Einstein--Maxwell theory \cite{Israel,Carter}
or the Einstein--massless scalar field theory \cite{Chase}.  Wheeler's
conjecture that baryon and like numbers cannot be specified for a black hole
set off a longstanding trend in the search for new ``no--hair theorems''.
Thus Teitelboim as well as Hartle \cite{Teitelboim} proved that the
nonelectromagnetic force between two ``baryons'' or ``leptons'' resulting
from exchange of various force carriers would vanish if one of the particles
was allowed to approach a black hole horizon. A different approach which I
developed \cite{BekPRL} was to show that classical massive scalar or vector
fields cannot be supported at all by a stationary black hole exterior, making
it impossible to infer any information about their sources in the black hole
interior. Although those ``no hair'' results and later generalizations
\cite{Adler} obviously supported Wheeler's original conjecture, they were
manifestly limited in scope.

The discovery in the early 1990's of the nonabelian gauge field bearing black
holes \cite{gauge} occasioned great surprise.  It should not have done so !
An old argument \cite{Bekthesis} reminds us that due to the gauge invariance
of electrodynamics, the Coulomb potential can propagate instantaneously in an
appropriate gauge.  Thus the potential from charges in a black hole can
cross the event horizon outward (going out of the tangent light cone), and so
convey to a distant observer the value of the hole's electric charge.  Gauge
invariance of a nonabelian gauge theories should likewise allow one or more
of the gauge field components generated by sources in a black hole to
``escape'' from it.  Thus gauge fields around a black hole may be possible
in every gauge theory.  Early on Yasskin \cite{Yasskin} exhibited some
such solutions in nonabelian gauge theories.

But not every field is allowed around a stationary black hole.  As we shall
see, scalar fields, optionally in partnership with Abelian gauge fields, are
mostly excluded.  Let us write the generic action for Abelian scalar field
theory,
\begin{equation}
S=S_{\rm EHM} - {1\over 2}\int{\left[
		   D_\alpha \psi {(D^\alpha\psi)}^*
	 	  + \xi R\psi^*\psi + V(\psi^*\psi)
				\right] \sqrt{-g}\, d^4x}.
    \label{action}
\end{equation}
Here $S_{\rm EHM}$ stands for the Einstein--Hilbert--Maxwell
action, $D_\alpha\equiv \partial_\alpha - \imath e A_\alpha$ with $A_\alpha$
the electromagnetic 4--potential, $R$ is the Ricci curvature scalar and $\xi$
measures the strength of the nonminimal gravitational coupling.  We have
already mentioned an early no scalar--hair theorem \cite{BekPRL}; it covered
only the case
$\xi=0$, $e=0$ (and therefore real $\psi$) and a potential such that $V'\geq 0$.
The all important case of the Mexican--hat potential $V(a^2)\propto
(a^2-a_0^2)^2$, still for $\xi=0$, was first handled by Adler and Pearson
\cite{Adler}.

For the conformal coupling case $\xi=1/6$, Bocharova,  Bronnikov and Melnikov
(BBM) \cite{Bro70} found a maverick black hole.  Its metric, electric and
scalar  fields are
\begin{eqnarray}	  ds^2 &=& -(1-M/r)^2 dt^2 + (1-M/r)^{-2} dr^2 + r^2
d\Omega^2\nonumber
\\
 F_{\mu\nu} &=& Q\,r^{-2}\,(\delta_\mu^r \delta_\nu^t - \delta_\mu^t
\delta_\nu^r);\qquad Q < M\label{BBM}
\\
\psi &=& \pm (3/4\pi)(M^2-Q^2)^{1/2}(r-M)^{-1}\nonumber
\end{eqnarray}
Published in Moscow's State University bulletin, the solution was at first
unknown in the West. I rediscovered it \cite{Bekscalar} and first showed that
the apparent singularity of $\psi$ at the horizon $r=M$ has no deleterious
consequences: a particle, even one coupled to $\psi$, encounters no infinite
tidal forces upon approaching the horizon. The mass $M$ and charge $Q$ of the
BBM black hole are independent parameters, in contrast with the situation for
the extremal RN black hole which has like metric.  But the presence of the
scalar field is reflected merely in a sign degree of freedom, a dichotomic
parameter, not a continuous one.  In one of his last papers before his tragic
death, Xanthopoulos in cooperation with Zannias, and then Zannias alone
\cite{Xanthopoulos} proved that there is no nonextremal extension of the BBM
black hole which might have made the extra parameter continuous.
The situation is not materially changed by the addition of magnetic charge
\cite{Virbhadra}.  Thus the BBM black hole has hair, but only in the guise of
a dichotomic parameter, very scanty hair indeed !  Add to this the fact that
the hole is known to be linearly unstable \cite{Kir}, and it is clear that the
``no scalar hair'' idea is hardly compromised by its existence.

The black hole pierced by a Nielsen--Olesen vortex or cosmic string \cite{Achu} is
another example of possible scalar hair.  No closed analytic solution exist for it;
claims for existence of this configuration by Achucarro, Gregory and Kuijken
rest on perturbative analytic and numerical arguments.  The authors point to
a distinction between the internal mass of the black hole and the
gravitating mass, and suggest that the presence of the string allows external
observers to get information about the last stage of formation of the hybrid.
Nevertheless, the solution is not globally asymptotically flat (the string
causes the usual conical singularity).  We know from Zel'dovich and Novikov
\cite{Zel} that once asymptotic flatness is given up, nonstandard black hole
geometries are allowed, but that only means that a black hole is affected by
far away masses, not that it has intrinsic new parameters.  It may thus well
be that in the black hole pierced by a vortex, the whole is no greater than
the sum of the parts.

Based on the foregoing fragmentary evidence, and theorems to be outlined, I
venture to make the following ``no scalar hair'' conjecture: {\it The only
asymptotically flat, spherically symmetric and truly static black
hole solutions of action (\ref{action}) are the RN and
generalized BBM families\/}.  The claim is that in Abelian scalar theory
the usual ``no hair'' idea works, at least for spherical configurations.
At this stage the spherical symmetry restriction is introduced to exclude
the ``black hole pierced by a vortex'' from discussion.  I believe it should
be possible to extend the conjecture to all static solutions with the
spherical symmetry assumption dropped in exchange for a topological one.

How to prove the conjecture ? Without restricting the black
hole configurations one  regards as physical by some kind of energy condition,
the conjecture cannot be proved.  For instance, consider the action
(\ref{action}) with $e=0$ (neutral field),
$\xi=1/6$, $V=0$,  but take the sign of the scalar action opposite that
shown.  Of course the field now bears negative energy.  We can get a new
black hole solution by replacing in the expression for $\psi$ in (\ref{BBM})
$(M^2-Q^2)^{1/2}\Rightarrow (Q^2-M^2)^{1/2}$ so that the new solution is valid
only for $Q>M$.  The result is a ``hairy'' black hole,
but one that could never form by collapse, for the Coulomb repulsion would
surely prevail over gravity and arrest the infall.  Giving up on positive
energy thus allows new black holes, but they are not interesting.

In the ongoing effort to establish the conjecture, the following theorem for a
real scalar field $\psi$ is very useful.  Take the action as
\begin{equation}
S=S_{\rm EHM}-\int {\cal E}({\cal I}, \psi)\sqrt{-g}\, d^4x;\qquad
{\cal I}\equiv g^{\alpha\beta}\psi,_\alpha\psi,_\beta
\end{equation}
with $\partial{\cal E}/\partial{\cal I} > 0$ and ${\cal E}>0$ for positive
argument.   With such stipulation any static $\psi$ configuration bears
positive energy\cite{Bek95}.  The theorem, proved by appealing to the
conservation of energy--momentum and Einstein's equation, states that the only
static spherically symmetric black hole solution is the RN one with no scalar
field
\cite{Bek95,MB}.  The theorem admits generalization to many scalar fields
\cite{Bek95} and special cases of it were known previously
\cite{Heusler}.

Now rewrite action (\ref{action}) for a neutral $\psi$ and positive
definite potential in terms of the metric \cite{Saa,MB}
\begin{equation}
\bar{g}_{\mu\nu}\equiv g_{\mu\nu}\Omega \equiv g_{\mu\nu}(1-8 \pi G \xi
|\psi|^2)
\end{equation}
to get
\begin{eqnarray}
S&=&S_{\rm EHM}[\bar{g}_{\mu\nu}]-\int{\left(
f\bar{g}^{\alpha\beta}\psi,_\alpha \psi,_\beta + \bar{V}
		\right)\sqrt{-\bar{g}}\,d^4x}\nonumber\\
	 f&\equiv& 1+ 48 \pi G \xi^2 \psi^2 (1-8 \pi G \xi \psi^2)^{-2}\\
	\bar{V}&\equiv& V(\psi^2)\cdot(1-8 \pi G \xi \psi^2)^{-2}\nonumber
 \end{eqnarray}
In the new theory black holes with a scalar field are ruled out by
the just mentioned theorem because ${\cal
E}=f\bar{g}^{\alpha\beta}\psi,_\alpha \psi,_\beta +  \bar{V}$
obviously obeys the conditions of positive energy, all this provided
the metric $\bar g_{\mu\nu}$ is regular and of like signature to
$g_{\mu\nu}$.  This is obviously so for $\xi\leq 0$ \cite{Saa,MB}.
For the harder case $\xi\geq 1/2$ Mayo and Bekenstein prove \cite{MB} that if
in the original geometry  $g_{\mu\nu}$ the Poynting vector of $\psi$'s
energy--momentum tensor is timelike (causality), then $1-8 \pi G \xi
\psi^2$ cannot change sign between the horizon ${\cal H}$ and infinity.  Now
asymptotically $1-8 \pi G \xi \psi^2 > 0$ in order that the effective
gravitational constant of the original theory be positive.  Hence if
instead of restricting attention to positive energy solutions, we
look only at solutions which obey a a causality restriction, the
metric  $\bar{g}_{\mu\nu}$ is regular and of appropriate signature
all the way from ${\cal H}$  to infinity, and neutral scalar hair is again
ruled out by the theorem.  In summary, {\it for positive definite potential
and for $\xi\leq 0$ {\rm or\ } $\xi\geq 1/2$, the black hole solutions of the
action (\ref{action}) for real $\psi$ and $e=0$ are restricted to the RN
family}.

Consider now a truly static complex $\psi$ governed by the
action(\ref{action}).  An early no scalar hair theorem was given by
Adler and Pearson \cite{Adler} for the special case $\xi=0$ with
$V(a^2)=(a^2-a_0^2)^2$.  More generally let us introduce the static
spherically symmetric metric
\begin{equation}
	ds^2 = - e^\nu dt^2 + e^\lambda dr^2 + r^2 (d\theta^2 + \sin^2\theta\,
d\phi^2)
\label{metric}
\end{equation}
Regardless of the matter content of the black hole exterior, the metric
coefficients near ${\cal H}$ must behave as \cite{MB}
\begin{equation} e^\nu \sim e^{-\lambda}\sim\cases{ r-r_H,&nonextremal;\cr
(r-r_H)^2, &extremal.\cr}
\label{asym}
\end{equation}
Writing out the Maxwell equation for the scalar potential
\begin{equation}
A_t,_{rr} + ({2\over r} - {\nu'+\lambda'\over 2}) A_t,_r
- 4 \pi e^2 |\psi|^2 e^\lambda A_t = 0
\label{At}
\end{equation}
we notice that there can be no maxima or minima of $A_t$ if that field
vanishes asymptotically \cite{MB}.  Hence $|A_t|$ grows monotonically from
$r=\infty$ to ${\cal H}$; it cannot, however, diverge on ${\cal H}$
because then a physical component of the electric field would blow up there
\cite{MB}.  As a consequence $A_{t,rr}$, if it diverges at all on ${\cal
H}$, can do so only softer than $(r-r_{\cal H})^{-1}$.   $|\psi|$ must then
vanish on ${\cal H}$, for otherwise the singularity in the last term of
Eq.(\ref{At}) could not be balanced.

The phase $\varphi$ of $\psi$ cannot depend on $r$ for then there would be a
radial current $\propto |\psi|^2 \varphi_{,r}$ which is incompatible with a
static situation.  The scalar equation is thus
\begin{equation}
|\psi|,_{rr} + ({2\over r} + {\nu'-\lambda'\over 2}) |\psi|,_r   -(\xi R +
V'-e^2 A_t^2 e^{-\nu}) e^\lambda |\psi|=0
\end{equation}
In light of the behavior (\ref{asym}), the finiteness of $A_t$ and of $R$ on
${\cal H}$, and an additional assumption that the potential is regular for
finite arguments, the generic asymptotic  solution of this equation near
${\cal H}$ is
\begin{equation}
|\psi|=B \sin\bigl( NeA_t(r_{\cal H})\ln(r-r_{\cal H}) + C\bigr)
\end{equation}
where $N$ is a combination of constants, and $B$ and $C$ are
integration constants.  Obviously $\psi$ does not vanish on ${\cal H}$ unless
it vanishes identically; the implication is that there is no solution except
the trivial one.  No assumption about the sign of $V$ or of $\xi$ need be made
here. Hence {\it for any regular potential and for any $\xi$, the stationary
black hole solutions of the action (\ref{action}) can have no charged scalar
hair, and are restricted to the RN family.\/}

To sum up, were one able to rule out neutral scalar hair based on the action
(\ref{action}) with positive definite potential for the ranges $0<\xi<1/6$
and $1/6<\xi<1/2$, then the ``no scalar hair'' conjecture would be
established.

This review is based on joint work with A. Mayo.


\begin{thebibliography}{99}
\setlength{\parskip}{0.32ex}


\bibitem{Wheeler} R. Ruffini and J. A. Wheeler, Physics Today {\bf 24}, 30
(1971).
\bibitem{gauge} M.S. Volkov and D.V. Gal'tsov, JETP Lett. {\bf 50}, 312
(1989); P.Bizon, Phys. Rev. Lett. {\bf 64}, 2844 (1990);  S. Droz, M. Heussler
and N. Straumann, Phys. Lett. {\bf B268}, 371 (1991); K-Y. Lee, V.P.
Nair and E. Weinberg, Phys. Rev. Lett. {\bf 68}, 1100 (1992); B. R. Greene,
S. D. Mathur and C. M. O'Neill, Phys. Rev. D{\bf 47}, 2242 (1993); T. Torii
and K. Maeda, Phys. Rev. D{\bf 48}, 1643 (1993).
\bibitem{Israel} W. Israel, Phys. Rev.  {\bf 164}, 1776 (1967) and Commun.
Math. Phys. {\bf 8}, 245 (1968).
\bibitem{Carter} B. Carter, Phys. Rev. Letters {\bf 26}, 331 (1971); R. M.
Wald, Phys. Rev. Letters {\bf 26}, 1653 (1971).
\bibitem{BekPT} J. D. Bekenstein, Physics Today {\bf 33}, 24 (1980).
\bibitem{Chase} J. E. Chase, Commun. Math. Phys. {\bf 19}, 276 (1970).
\bibitem{Teitelboim} C. Teitelboim, Lett.  Nuov. Cim. {\bf 3}, 326 and 397
(1972); J. Hartle,  Phys. Rev. D  {\bf 3}, 2938 (1971)
\bibitem{BekPRL} J. D. Bekenstein, Phys. Rev. Letters {\bf 28}, 452
(1972); Phys. Rev. D  {\bf 5}, 1239 and 2403  (1972).
\bibitem{Adler} S. A. Adler and R. P. Pearson,  Phys. Rev. D  {\bf 18},
2798  (1978).
\bibitem{Bekthesis} J. D. Bekenstein, Dissertation, Princeton University
(1972).
\bibitem{Yasskin}  P. Yasskin (1972), unpublished.
\bibitem{Bro70} N. Bocharova, K. Bronnikov and V. Melnikov, Vestn. Mosk.
Univ. Fiz. Astron. {\bf 6}, 706 (1970).
\bibitem{Bekscalar} J. D. Bekenstein, Ann. Phys. (NY)  {\bf 82}, 535
(1974);  Ann. Phys. (NY)  {\bf 91}, 72  (1975).
\bibitem{Xanthopoulos} B. C. Xanthopoulos and T. Zannias, J. Math. Phys.
{\bf 32}, 1875 (1991);  T. Zannias, J. Math. Phys. {\bf 36}, 6970 (1995).
\bibitem{Virbhadra} K. S. Virbhadra and J. C. Parikh,  Phys. Lett. B {\bf
331}, 302 (1994).
\bibitem{Kir}  K. A. Bronnikov and Yu. N. Kireyev, Phys. Lett. {\bf 67A},
95 (1978).
\bibitem{Achu}  A. Achucarro,  R. Gregory and K. Kuijken Phys. Rev. D  {\bf
52}, 5729  (1995).
\bibitem{Zel} Ya. B. Zel'dovich and I. D. Novikov, {\it Stars and
Relativity\/}  (University of Chicago Press, Chicago, 1971), p.138.
\bibitem{Bek95} J. D. Bekenstein, Phys. Rev. D  {\bf 51}, R6608 (1995).
\bibitem{MB} A. E. Mayo and J. D. Bekenstein, Phys. Rev. D  {\bf 54}, xxx
(1996).
\bibitem{Heusler} M. Heusler, J. Math. Phys. {\bf 33}, 3497 (1992);
Class. Quant.  Grav. {\bf 12}, 779 (1995); D. Sudarsky,  Class. Quant. Grav.
{\bf 12}, 579 (1995).

\end{thebibliography}
\end{document}